\documentclass{cpbtex}
\newcommand{\er}{\mathrm{e}}
\newcommand{\ir}{\mathrm{i}}
\newcommand{\dr}{\mathrm{d}}
\newcommand{\pp}{p}

\usepackage{bm}
\usepackage{multirow}
\usepackage{diagbox}
\usepackage{graphicx}
\usepackage{cmap}

\def\no{\nonumber}

\begin{document}


\title{Exact spectrum of the XX spin chain with constrained non-diagonal boundary fields}

\author{Changqing Liu$^{1}$, \ Xiaotian Xu$^{1,2,3,4}$\thanks{E-mail:xtxu@nwu.edu.cn}, \ and \ Xin Zhang$^{5}$\thanks{E-mail:xinzhang@iphy.ac.cn}\\
$^{1}${Institute of Modern Physics, Northwest University, Xi'an 710127, China}\\
$^{2}${Peng Huanwu Center for Fundamental Theory, Xi'an 710127, China}\\
$^{3}${Shaanxi Key Laboratory for Theoretical Physics Frontiers, Xi'an 710127, China}\\
$^{4}${Fundamental Discipline Research Center for Quantum Science and Technology}\\
{of Shaanxi Province, Xi'an 710127, China}\\
$^{5}${Beijing National Laboratory for Condensed Matter Physics, Institute of Physics,}\\
{Chinese Academy of Sciences, Beijing 100190, China}}

\date{\today}
\maketitle

\begin{abstract}
We study the exact spectrum of the XX spin chain with constrained non-diagonal boundary fields, which can be analyzed by solving the associated Bethe Ansatz equations. In these equations, the number of Bethe roots has a definite parity, and all Bethe roots are located at the zeros of a unary function. We investigate the possible positions of the Bethe roots. Based on numerical observations, we analyze the Bethe root configurations for the ground state and the first excited state. Our results show that elementary excitations are characterized by the cooperative change of a pair of Bethe roots. Furthermore, we obtain an analytical expression for the ground state energy in the thermodynamic limit.

\end{abstract}


\section{Introduction}
The Heisenberg model \cite{heisenberg1928} plays a critical role in condensed matter and statistical physics. It not only captures a variety of essential physical mechanisms \cite{mourigal:2013,mikeska:2008}, but also provides a fundamental theoretical framework for understanding quantum ``phase'' transitions, spin-liquid behavior, and high-temperature superconductivity \cite{liu:2022gapless,broholm:2020liquids,jiang:2021super}. 

The Heisenberg model is one of the most well-studied integrable models \cite{Baxter1982,korepin1997quantum,wang2015off}. Bethe first solved the model and thereby established the Bethe Ansatz method \cite{bethe1931}. Since then, various analytical methods have been successfully applied to derive the exact solutions of the Heisenberg model under different integrable boundary conditions, such as Baxter's $T$-$Q$ relation \cite{Baxter1982,Baxter:2001,nepomechie2003,nepomechie2003completeness,Yang:2005tq}, the (modified) algebraic Bethe Ansatz \cite{takhtadzhan1979,korepin1997quantum,cao2003exact,belliard2013}, the thermodynamic Bethe Ansatz \cite{takahashi1999}, the off-diagonal Bethe Ansatz \cite{wang2015off,cao2013off,cao2013off1,cao2013off2}, the separation of variables \cite{niccoli2012,niccoli2013,Faldella:2014}, the modified coordinate Bethe Ansatz \cite{zhang2021phantom,Zhang:2021chiral}, and so on.

After the pioneering work of Sklyanin and Cherednik \cite{sklyanin1988,cherednik1984}, people realized that the integrability of the model can also be preserved in open systems. For both the isotropic and anisotropic spin-$1/2$ Heisenberg chain in one dimension, quantum integrability allows one to add magnetic fields at the two boundaries. The boundary fields play a crucial role in determining ground-state properties and low-energy excitations \cite{Li:2014,Sun:2019,Qiao:2021,dong2023exact,kozlowski:2012surface,pozsgay:2018exact,kapustin:1996surface} and also make the construction of the exact solutions more challenging than that in the periodic case. In the generic case, the boundary fields break the $U(1)$ symmetry of the system. As a consequence, one cannot use the conventional $T$-$Q$ relation and Bethe Ansatz equations (BAEs) to parametrize the exact solutions of the system. Instead, one can add an inhomogeneous term to the $T$-$Q$ relation, which results in a significantly different form of the BAEs \cite{wang2015off,cao2013off1,cao2013off2}. Even in the case with $U(1)$ symmetry, the BAEs have a symmetric form; however, they cannot be solved exactly due to the coupling structure of the Bethe roots and the introduction of boundary parameters \cite{korepin1997quantum,takahashi1999}. To date, the study of the spectrum of the Heisenberg chain with non-diagonal boundary fields has mainly focused on two aspects: (i) the small-size case, where one can use numerical methods to diagonalize the Hamiltonian or solve the BAEs \cite{Jiang:2025}; and (ii) the thermodynamic limit, where the properties of the ground state and low-lying excited states can be derived analytically \cite{cao2003exact,Li:2014,Sun:2019,Qiao:2021,murgan:2006boundary}. For intermediate-scale systems, the exact spectrum of the model has still not been systematically characterized.

It has been shown that for anisotropic Heisenberg chains with non-diagonal boundary fields, homogeneous $T$-$Q$ relations and conventional Bethe Ansatz equations can still be constructed under certain constraints \cite{cao2003exact,nepomechie2003,Yang:2005tq}. In such cases, although the $U(1)$ symmetry is broken, the system may still possess other symmetries \cite{Chernyak:2022}. Two major differences between these non-trivial ``degenerate'' points and the $U(1)$-symmetric case are: (i) the number of Bethe roots is fixed, (ii) additional boundary parameters are introduced. 

Motivated by these observations, in this paper we investigate the XX spin-chain model with such constrained non-diagonal boundary fields, as given by Eqs. \eqref{xx Hamiltonian} and \eqref{Constraint:XX}. Specifically, the constraint in \eqref{Constraint:XX} is characterized by an integer $M$, which denotes the total number of Bethe roots and has a fixed parity.

Another property of the model studied in this paper is that the corresponding BAEs do not involve coupling among Bethe roots. This implies that all Bethe roots are located at the zeros of an auxiliary unary function. Consequently, we are able to study the exact spectrum of the model for finite-size systems, for intermediate-scale systems (ranging up to hundreds or even thousands of lattice sites), as well as in the thermodynamic limit.

It is worth noting that, with constrained non-diagonal boundary conditions, the elementary excitation patterns vary with the system parameters and differ from those in the diagonal case. In systems with diagonal boundary fields, excitations are typically associated with the addition or removal of a single Bethe root \cite{lieb1961two,biegel:2004spectrum}, whereas under these constraints they are characterized by the addition, removal, or exchange of a pair of Bethe roots.

The paper is organized as follows. Section \ref{sec:Ham} introduces the open XX spin chain with constrained non-diagonal boundary fields, presenting both the Bethe Ansatz equations and the energy expression. In Section \ref{Zero distribution}, we analyze the possible distribution locations of Bethe roots. Section \ref{sec:ground:state} is about the distributions of Bethe roots in the the ground state and the first excited state. An analytical expression for the ground state energy in the thermodynamic limit is derived in Section \ref{sec:therm}. In Section \ref{sec:diag}, we discuss the exact spectrum of the XX model with diagonal boundary fields. We conclude in Section \ref{sec:Conclusion}. The derivation of the Bethe Ansatz equations for the XX spin chain is provided in Appendix~\ref{App:A}.

\section{The XX spin chain with with constrained non-diagonal boundary fields}\label{sec:Ham}
Let us first introduce the XXZ spin chain with generic boundary magnetic fields, whose Hamiltonian reads
\begin{align}\label{Hamiltonian: XXZ}
H_{\rm XXZ}&=\sum_{j=1}^{N-1} \left(\sigma_j^x \sigma_{j+1}^x + \sigma_j^y \sigma_{j+1}^y +\cosh\eta \,\sigma_j^z\sigma_{j+1}^z\right) +\vec{h}_{\rm l}\cdot\vec{\sigma}_1+\vec{h}_{\rm r}\cdot\vec{\sigma}_N,\notag \\
\vec{h}_{\rm l}&=\frac{\sinh\eta}{\sinh {\bar\alpha}_- \cosh {\bar\beta}_-} 
\left( \cosh {\bar\theta}_-,\, {\rm i}\sinh {\bar\theta}_-,\, \cosh {\bar\alpha}_- \sinh {\bar\beta}_- \right), \notag \\
\vec{h}_{\rm r}&=\frac{\sinh\eta}{\sinh \bar\alpha_+ \cosh \bar\beta_+} 
\left( \cosh \bar\theta_+,\,{\rm i}\sinh \bar\theta_+,\,- \cosh {\bar\alpha}_+ \sinh {\bar\beta}_+ \right).
\end{align}
This model is one of the most well-known quantum integrable systems. In the generic case, the non-diagonal boundary fields break the $U(1)$ symmetry of the system, rendering conventional Bethe Ansatz approaches inapplicable \cite{wang2015off}.

However, under the following constraint
\begin{eqnarray}
(N-1-2M)\eta=\bar\alpha_-+\bar\alpha_++\bar\beta_-+\bar\beta_+\pm(\bar\theta_--\bar\theta_+)+2{\rm i} m\pi,\quad M,m\in\mathbb{Z},\label{Constraint:XXZ}
\end{eqnarray}
the exact solutions of the model, including its eigenvalues and eigenstates, can be constructed via either the modified algebraic Bethe Ansatz or the modified coordinate Bethe Ansatz method \cite{cao2003exact,nepomechie2003,Zhang:2021chiral}.

In this paper, we aim to study an open XX model ($\eta={\rm i}\pi/2$) under Eq. (\ref{Constraint:XXZ}). The Hamiltonian reads 
\begin{align}\label{xx Hamiltonian}
H_{\rm XX} &=  \sum_{j=1}^{N-1} \left( \sigma_j^x \sigma_{j+1}^x + \sigma_j^y \sigma_{j+1}^y \right) +\vec{h}_{\rm l}\cdot\vec{\sigma}_1+\vec{h}_{\rm r}\cdot\vec{\sigma}_N,\notag \\
\vec{h}_{\rm l}&=\frac{1}{\sin \alpha_- \cosh \beta} 
\left( \cos \theta_- ,\, - \sin \theta_-,\, \cos \alpha_- \sinh \beta \right), \notag \\
\vec{h}_{\rm r}&=\frac{1}{\sin \alpha_+ \cosh \beta} 
\left( \cos \theta_+ ,\,- \sin \theta_+, \,\cos \alpha_+ \sinh \beta \right),
\end{align}
where 
\begin{eqnarray}
 (N-2M-1)\pi/2=\alpha_-+\alpha_+\pm(\theta_--\theta_+)+2m\pi,\quad m,M\in\mathbb{Z}.\label{Constraint:XX}
\end{eqnarray}
To ensure that the Hamiltonian $H_{\rm XX}$ is Hermitian, we impose $\bar\beta_- = -\bar\beta_+ = \beta$, and take $\alpha_\pm$ and $\theta_\pm$ to be real numbers. When we shift $M$ by 2, Eq. (\ref{Constraint:XX}) remains valid. Therefore, the integer $M$ in Eq. (\ref{Constraint:XX}) is restricted to a specific parity-it can only take odd or even values.

We can characterize the two boundary fields using their amplitudes $|\vec{h}_{\rm l,r}|$, polar angles $\varphi_{\rm l,r}$, and azimuthal angles $\theta_{\rm l,r}$, specifically as 
\begin{eqnarray}
\vec{h}_{\rm l,r}=|\vec{h}_{\rm l,r}|(\cos\theta_{\rm l,r}\sin\varphi_{\rm l,r},\,\sin\theta_{\rm l,r}\sin\varphi_{\rm l,r},\,\cos\varphi_{\rm l,r}).
\end{eqnarray}
While the amplitudes and polar angles can be free parameters, 
the azimuthal angles of the two boundary fields should maintain a fixed difference
\begin{eqnarray}
\theta_{\rm l}-\theta_{\rm r}\equiv \theta_+-\theta_-=\pm\vartheta_{N,M},\quad \vartheta_{N,M}=(N-2M-1)\pi/2-\alpha_--\alpha_+,
\end{eqnarray}
which, however, is a complex function of the amplitudes $|\vec{h}_{\rm l,r}|$ 
and polar angles $\varphi_{\rm l,r}$. 

\paragraph{Bethe Ansatz equations} The exact solution of the model given in Eqs. \eqref{xx Hamiltonian} and \eqref{Constraint:XX} can be parameterized by the following Bethe Ansatz equations 
\begin{align}
&\left[\frac{\sinh(\mu_j+\frac{\ir\pi}{4})}{\sinh(\mu_j-\frac{\ir\pi}{4})}\right]^{2N}\frac{\sinh(\mu_j-\ir\alpha_--\frac{\ir\pi}{4}) \sinh(\mu_j-\ir\alpha_+-\frac{\ir\pi}{4}) }{\sinh(\mu_j+\ir\alpha_-+\frac{\ir\pi}{4}) \sinh(\mu_j+\ir\alpha_++\frac{\ir\pi}{4}) }\nonumber\\
&\times \frac{\cosh(\mu_j-\beta-\frac{\ir\pi}{4}) \cosh(\mu_j+\beta-\frac{\ir\pi}{4})}{\cosh(\mu_j-\beta+\frac{\ir\pi}{4}) \cosh(\mu_j+\beta+\frac{\ir\pi}{4})}=1,\qquad j=1,\ldots,M,
\label{BAEs}
\end{align}
where $M$ is determined in \eqref{Constraint:XX}.
The energy in terms of the Bethe roots (or the rapidities) $\{\mu_1,\dots,\mu_M\}$ is given by
\begin{align}
&\mathcal{E}(\bm{\mu})=\sum_{j=1}^M \varepsilon(\mu_j)+E_0,\quad 
E_0=-\cot \alpha_{-}-\cot \alpha_{+},\no\\
&\varepsilon(u)=-\frac{4}{\cosh(2u)},\quad \bm{\mu}=\{\mu_1,\ldots,\mu_M\}.\label{Energy expressions 1}
\end{align}
Obviously, the boundary-related parameters in the BAEs \eqref{BAEs} are just functions of the amplitude and the $z$-component of the boundary fields. From Eqs.~\eqref{Constraint:XX}, \eqref{BAEs}, and \eqref{Energy expressions 1}, we see that the energy spectrum of the model depends on the difference $\theta_{\rm l}-\theta_{\rm{r}}=\theta_+-\theta_-$, but not on the individual parameters $\theta_{\rm{l}}$ and $\theta_{\rm{r}}$. In other words, systems with a fixed $\theta_{\mathrm{l}} - \theta_{\mathrm{r}}$ but different $\theta_{\mathrm{l}}$ and $\theta_{\mathrm{r}}$ share the same energy spectrum.

Define the following parameters
\begin{eqnarray}
w_\pm=\cot\alpha_\pm,\quad g=\tanh^2\beta,
\end{eqnarray}
which are related to the boundary fields as follows
\begin{eqnarray}
&|\vec{h}_{\rm l}|^2=w_-^2-g+1,\quad (\vec{h}_{\rm l})^2_z=w_-^2\,g,\\
&|\vec{h}_{\rm r}|^2=w_+^2-g+1,\quad (\vec{h}_{\rm r})^2_z=w_+^2\,g.
\end{eqnarray}
The BAEs in \eqref{BAEs} can be rewritten in terms of the quasi-momentum
\begin{align}
p_j=-\ir \ln\left[\frac{\sinh(\mu_j+\frac{\ir\pi}{4})}{\sinh(\mu_j-\frac{\ir\pi}{4})}\right],
\end{align}
as follows
\begin{align}
&f(\er^{\ir p_j})=0, 
\qquad j=1, \ldots, M,\label{BAEs(alter)}\\
&f(u) =u^{2N}( g +u^2) \prod_{\sigma =w_\pm}{( \sigma-u ) -}( u^2 g +1 ) \prod_{\sigma =w_\pm}{( 1-u\sigma )}.\label{auxiliary function}
\end{align}
The energy can also be parameterized by $\bm{p}=\{p_1,\ldots,p_M\}$
\begin{eqnarray}\label{Energy expressions}
E(\bm{p})=\sum_{j=1}^M\epsilon(p_j)-w_+-w_-,\quad \epsilon(x)=4\cos x.
\end{eqnarray}

\paragraph{Completeness of BAEs}
The Bethe roots \(\bm{p}=\{p_1,\dots,p_M\}\) must satisfy the selection rule 
\begin{align}
\er^{2\ir p_j}\neq 1,\quad \er^{\ir (p_j - p_k)} \neq 1, \quad \er^{\ir (p_j + p_k)} \neq 1,\quad j\neq k. \label{selection:rule}
\end{align}
It should be also noted that $p_j$ and $-p_j$ are equivalent and $\er^{\ir p_j}=\pm 1$ are trivial solutions. Since all Bethe roots are zeros of the function $f(\er^{\ir  {x}})$, solving the Bethe Ansatz equations can reduce to analyzing the independent non-trivial zeros of the auxiliary function \(f(u)\).

The function $f(u)$ is a rational polynomial of degree $2N+4$, and its zeros come in reciprocal pairs due to the anti-reciprocal symmetry
\[
f(u^{-1}) = -u^{-2N-4} f(u).
\]
We see that \(f(u)\) possesses \(N+1\) independent non-trivial zeros, denoted as $\{u_1,\ldots,u_{N+1}\}$, which can be restricted to the closed unit disc $$\{u\in\mathbb{C}||u|\leq 1\},$$
and a physical solution of the BAEs \eqref{BAEs(alter)} should thus satisfy
\begin{eqnarray}\label{e-u}
\{\mathrm{e}^{\mathrm{i}p_1},\ldots,\mathrm{e}^{\mathrm{i}p_M}\}\subseteq \{u_1,\ldots,u_{N+1}\},
\end{eqnarray}
or, equivalently, 
\begin{align}
\{p_1,\ldots,p_M\}\subseteq \{-\ir\ln(u_1),\ldots,-\ir\ln (u_{N+1})\}.
\end{align}

Equation \eqref{e-u} implies that the integer $M$
in \eqref{Constraint:XX} must satisfy $0\le M\le N+1$ and can be either even or odd.
For a fixed $M$, the BAEs \eqref{BAEs(alter)} have 
$\binom{N+1}{M}$
independent physical solutions. Due to the identity
\begin{eqnarray}\label{odd and even}
\sum_{{\rm odd}\,\, k}\binom{N+1}{k}=\sum_{{\rm even}\,\, k}\binom{N+1}{k}=2^N,
\end{eqnarray}
all the energy spectrum of the Hamiltonian can be obtained.

\paragraph{Remark.}
\emph{Unlike the case of diagonal boundary fields, the total number of the Bethe roots $M$ is not tied to the magnon number and may even exceed the system length $N$.}

For the XX chain with generic boundary field, the corresponding BAEs (see Eq. \eqref{BAE:generic:xx}) contain an inhomogeneous term and all the Bethe roots are entangled with each other. This greatly increases the difficulty of solving BAEs.

\section{Possible Configurations of the Bethe Roots}\label{Zero distribution}

Most zeros of $f(u)$ lie on the unit circle, corresponding to real quasi-momentum. In contrast, some real zeros may appear in the interval $(-1,1)$ and give rise to imaginary quasi-momentum (${\rm Re}(p)=0,\pi$).
Depending on the pattern of the zeros, the ``phase'' diagram can be divided into several regimes, as summarized in Fig. \ref{Zero distribution of auxiliary function}.

Noticing the following relations
\begin{align}
f(u)|_{w_\pm \to w_\mp}&=f(u),\quad \{u_1,\ldots,u_{N+1}\}_{w_\pm \to w_\mp}=\{u_1,\ldots,u_{N+1}\}, \\
f(u)|_{w_\pm \to -w_\pm}&=f(-u),\quad \{u_1,\ldots,u_{N+1}\}_{w_\pm \to -w_\pm}=\{-u_1,\ldots,-u_{N+1}\},\label{condition w}
\end{align}
the ``phase'' diagram possesses certain symmetry, see Fig. \ref{Zero distribution of auxiliary function}.
\begin{center}
\refstepcounter{figure}
\includegraphics{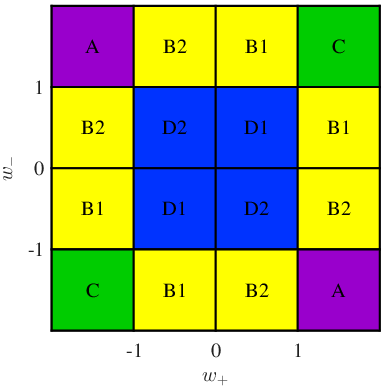}\\[5pt]  
\parbox[c]{15.0cm}{\footnotesize{\bf Fig.~1.}   Configuration of zeros of the auxiliary function $f(u)$ in different regions of the boundary parameters $w_\pm$. The zero configurations in regions B1(D1) and B2(D2) are quite similar, yet no direct relation exists between them.}
\label{Zero distribution of auxiliary function}
\end{center}

\subsection{Finite \texorpdfstring{$N$}{N} case}
Let us start with finite-size systems. First, the properties of $f(u)$ at some specific points can be analyzed
\begin{align}
f(\pm 1)&=0,\qquad f(0)=-1, \\
f(w_\pm)&=-(w_\pm^{2}g+1)(1-w_\pm^2)(1-w_+w_-),\label{fwpm1}\\
f(w_\pm^{-1})&=w_\pm^{-2N-4}\left(1+g w_\pm^{2} \right) \left( w_\pm^{2}-1 \right) \left( w_+w_--1 \right), \label{fwpm2}\\
f'(1) &=2(g+1)(w_+-1)(w_--1) (N-\bar a),\\
f'(-1)&=2(g+1)(w_++1)(w_-+1) (\bar b-N),  
\end{align}
where
\begin{align}
\bar a &=\frac{g-1}{g +1}+\frac{w_+w_--1}{(w_--1) (w_+-1)},\label{Differentiate a}\\
\bar b &=\frac{g-1}{g +1}+\frac{w_+w_--1}{(w_-+1) (w_++1)}.\label{Differentiate b}
\end{align}

The analysis procedure is the same for regions of the same color. In the following, we take some selected regions as examples.

\paragraph{Region A ($w_+>1,\,w_-<-1$)}

In this region, since $w_+w_-<-1$, we can prove that
\begin{eqnarray}
    f(w_+^{-1}),f(w_-^{-1})<0.
\end{eqnarray}
If $f'( 1 )<0$ and  $f'( -1 )>0$, i.e.:
\begin{eqnarray}\label{condition:1}
N>{\rm max}\{\bar a,\,\bar b\}.
\end{eqnarray}
The auxiliary function possesses two independent real zeros, with the positive real zero located within interval $(w_+^{-1},1) $, and the negative real zero situated within interval $(-1,w_-^{-1}) $. The remaining $N-1$ zeros are in the unit circle. As shown in Figure \ref{Fig: Zero distribution}(a). 
If $f'(1)<0$, $f'(-1)<0$ or $f'(1)>0$, $f'(-1)>0$, i.e.:
\begin{eqnarray}\label{condition:2}
    {\rm min}\{\bar a,\,\bar b\}<N<{\rm max}\{\bar a,\,\bar b\},   
\end{eqnarray}
the auxiliary function possesses one real zero located in the interval
$(w_+^{-1},1) $ or $(-1,w_-^{-1}) $, while the remaining $N$ zeros lie on the unit circle. As shown in Figure \ref{Fig: Zero distribution}(b).

Once $f'(1)>0$, $f'(-1)<0$, i.e.:
\begin{eqnarray}\label{condition:3}
N<{\rm min}\{\bar a,\,\bar b\}, 
\end{eqnarray}
all zeros of the auxiliary function lie on the unit circle.

\paragraph{Region B ($w_+>1>w_->0$)}

When $w_+w_->1$ ($w_->w_+^{-1}$), we find that $$f(w_+^{-1}),f(w_-) >0,$$
indicating that the auxiliary function $f(u)$ has a positive real zero in the interval $(0,w_+^{-1})$, while the remaining $N$ zeros are in the unit circle. As shown in Figure \ref{Fig: Zero distribution}(c).

When $w_+w_-<1$, we should examine the derivative $f'(1) $ of the auxiliary function $f(u)$ at $u = 1$. If $f'(1)<0$, (i.e., $N>\bar a$), the auxiliary function possesses a positive real zero within the interval $(w_+^{-1},1) $. The remaining $N$ zeros all lie on the unit circle, as shown in Figure \ref{Fig: Zero distribution}(d). If $f'(1)>0$ (i.e., $N<\bar a$), all zeros of the auxiliary function are thought to lie on the unit circle. Analogously, one can analyze the pattern of zeros in region B2.

\paragraph{Region C ($w_+>w_->1$)}
In this region, one can find
\begin{eqnarray}
f(w_+^{-1}),f(w_-^{-1}) >0.
\end{eqnarray}
The auxiliary function $f(u)$, therefore, admits one positive real zero within the interval $(0,w_+^{-1})$, as shown in Figure \ref{Fig: Zero distribution}(e).

Furthermore, if $f'(1)>0$ (i.e., $N>\bar a$), the auxiliary function $f(u)$ possesses another positive real zero at $(w_-^{-1},1) $, while the remaining $N-1$ zeros all lie on the unit circle, as shown in Figure \ref{Fig: Zero distribution}(f).

\paragraph{Region D ($0<w_-,w_+<1$)}
In this region, one can easily derive
\begin{eqnarray}
f(w_+),f(w_-)<0.
\end{eqnarray}
By substituting the boundary parameters into Eqs. \eqref{Differentiate a} and \eqref{Differentiate b}, one obtains:
\begin{eqnarray}
\bar a,\bar b<0,
\end{eqnarray}
which leads to $f'(1 )>0$, $f'(-1)<0$ for arbitrary $N$.
Consequently, all zero of the auxiliary function $f(u)$ may lie on the unit circle. Analogously, one can analyze the zero distribution patterns in region D2.

\begin{center}
\refstepcounter{figure}
\includegraphics[width=6.5cm]{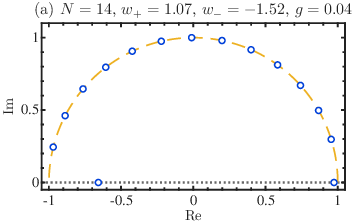}\qquad
\includegraphics[width=6.5cm]{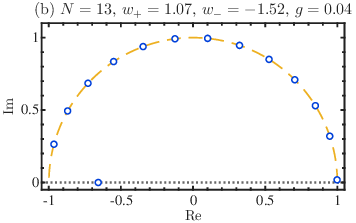}\qquad
\includegraphics[width=6.5cm]{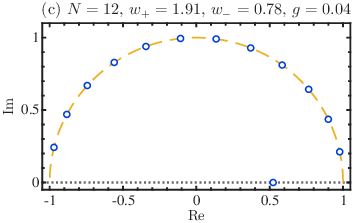}\qquad
\includegraphics[width=6.5cm]{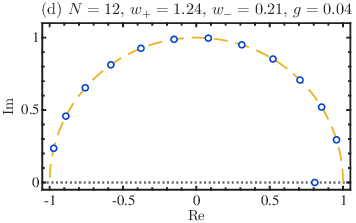}\qquad
\includegraphics[width=6.5cm]{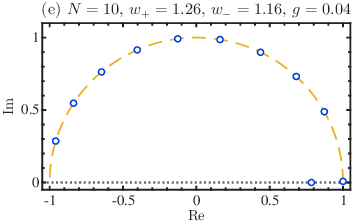}\qquad
\includegraphics[width=6.5cm]{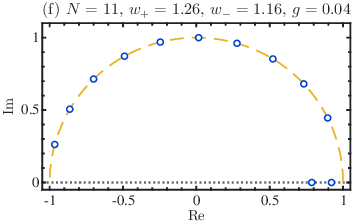}\\[8pt]
\parbox[c]{15.0cm}{\footnotesize{\bf Fig.~2.}   Distributions of $\{u_1,\ldots,u_{N+1}\}$ for different boundary parameters. The blue circles denote $\{u_1,\ldots,u_{N+1}\}$, the gray dashed line corresponds to ${\rm Im}(u)=0$, and the yellow dashed line represents the semicircle of unit radius. Panels (a) and (b) correspond to region A.  The real zeros undergo qualitative changes between $N=13$ and $N=14$. Panels (c) and (d) correspond to region B. The real zeros in the panels (c) and (d) are 0.5235 and 0.8075, respectively, located in the intervals $(0,w_+^{-1})$ and $(w_+^{-1},1)$. Panels (e) and (f) correspond to region C and resemble panels (a) and (b), except that the zeros on the real axis in (e) and (f) have the same sign (both positive or both negative).}
\label{Fig: Zero distribution}
\end{center}

\bigskip

For finite systems sizes, the corresponding zero distributions in each region are shown in Table \ref{Tab:Number of real zeros in each region}.
\begin{center}
\refstepcounter{table}
{\footnotesize{\bf Table 1.} Configurations of real $u_j$ at finite system size in different regions.\\
\vspace{2mm}
\begin{tabular}{|c|c|c|}
\hline
{Region}      & {number of real zeros}                         & {constraint} \\\hline
\multirow{3}{*}{A}
 & two  & $N > \max\{\bar{a},\bar{b}\}$ \\ \cline{2-3}
 & one  & $\min\{\bar{a},\bar{b}\} < N < \max\{\bar{a},\bar{b}\}$ \\ \cline{2-3}
 & zero & $ N < \min\{\bar{a},\bar{b}\}$ \\ \hline
\multirow{2}{*}{B}
 & one  & $N > \max\{\bar{a},\bar{b}\}$ \\ \cline{2-3}
 & zero & $  N < \min\{\bar{a},\bar{b}\}$ \\ \hline
\multirow{2}{*}{C}
 & two  & $N > \max\{\bar{a},\bar{b}\}$ \\ \cline{2-3}
 & one  & $\min\{\bar{a},\bar{b}\} < N < \max\{\bar{a},\bar{b}\}$ \\ \hline
\multirow{1}{*}{D}
 & zero & $N\in \mathbb{N}_+$ \\ 
\hline
\end{tabular}}
\label{Tab:Number of real zeros in each region}
\end{center}

\subsection{Large \texorpdfstring{$N$}{N} case}

When $|u_j|<1$ and $N$ is large enough, we have $u_j^{2N}\approx0$. Therefore,
\begin{eqnarray}
0=f(u_j)\approx  -(u_j^2g +1 ) (1-u_jw_-)(1-u_jw_+).
\end{eqnarray}
To satisfy this relation, we need either $u_j\approx w_-^{-1}$ or $u_j\approx w_+^{-1}$. 
Therefore, we can conclude the distribution of real zeros for a large $N$
\begin{align}
\begin{aligned}
&(1):\quad \mbox{when $|w_+|,|w_-|<1:\quad  $none real zeros}, \\
&(2):\quad \mbox{when $|w_\alpha|<1<|w_\beta|:\quad  $one real zero $u_j\approx w_\beta^{-1}$},\\
&(3):\quad \mbox{when $1<|w_+|,|w_-|:\quad  $two real zeros $u_j\approx w_+^{-1},w_-^{-1}$}.
\end{aligned}
\end{align}
It can be observed that $u_j$ tends to $w_\alpha^{-1}$ (with $|w_{\alpha}|>1$) exponentially with increasing system size; see Fig. \ref{u_j}.

\begin{center}
\refstepcounter{figure}
\includegraphics[width=0.5\textwidth]{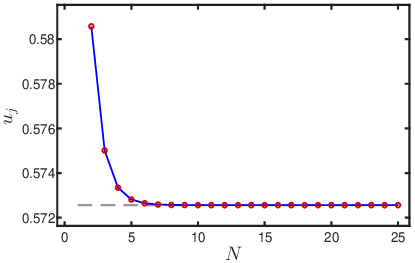}\\[5pt]  
\parbox[c]{15.0cm}{\footnotesize{\bf Fig.~3.}  Variation of the real zero with the system size $N$ for the boundary parameters $g=0.04$, $w_+=1.75$, and $w_-=0.36$. The red circles denote the numerical values of the zeros, while the blue solid line indicates their evolution trend as $N$ increases. The gray auxiliary line corresponds to $u_j = w_+^{-1}$.}
\label{u_j}
\end{center}

Real and imaginary quasi-momentum $p$ have different bare energies, and the imaginary $p$ plays a critical role when studying a given energy level, such as the ground state and the first excited state (see Section \ref{sec:ground:state} for more details).
In addition, imaginary quasi-momentum $p$ may lead to physical phenomena that are distinctly different from those of real $p$. In Figure \ref{Fig:Mag}, we consider a specific case of a single Bethe root as an example. For eigenstates corresponding to real $p$, the quantity $\langle\sigma_n^z\rangle$ exhibits nonlocal behavior and oscillates with $n$. In contrast, an imaginary $p$ causes this quantity to decay exponentially with $n$ and become localized near one boundary.

\begin{center}
\refstepcounter{figure}
\includegraphics[width=0.55\textwidth]{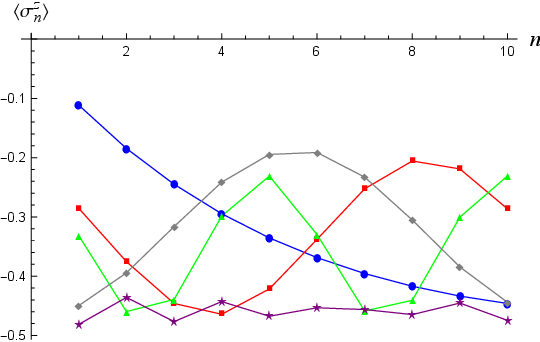}\\[5pt]  
\parbox[c]{15.0cm}{\footnotesize{\bf Fig.~4.}  $\langle\sigma_n^z\rangle$ versus $n$ for certain eigenstates. Here $N=10$, $M=1$, $\beta =0.50$, $\{\alpha _-,\alpha _+,\theta _-\}=\{0.73,1.00,0.78\}$. The blue, red, green, gray, and purple dots represent the eigenstates corresponding to $\bm{p}=\{0.0926\ir\}$, $\bm{p}=\{0.3495\}$, $\bm{p}=\{0.6267\}$, $\bm{p}=\{2.8491\}$, and $\bm{p}=\{1.4501\}$, respectively.}
\label{Fig:Mag}
\end{center}

\section{Distributions of Bethe roots in the ground state and first excited state}\label{sec:ground:state}
\subsection{Ground state}
Since all Bethe roots are non-interacting, one can easily study the energy spectrum of the model, among which the ground state and the first excited states are of most concern.

Let us first study the contribution of a single Bethe root to the energy
\begin{align}
\begin{aligned}
&({\rm i}):\,\,0\leq \epsilon(x)\leq 4,\quad  x\in[0,\pi/2],\\
&({\rm ii}):\,\,-4\leq \epsilon(x)<0,\quad  x\in(\pi/2,\pi],\\
&({\rm iii}):\,\,\epsilon({\rm i} x)\geq 4,\quad x\in\mathbb{R},\\
&({\rm iv}):\,\,\epsilon(\pi+{\rm i} x)\leq -4, \quad x\in \mathbb{R}.
\end{aligned}\label{BR:energy}
\end{align}
To derive the Bethe roots for the ground state, we can proceed as follows:
\begin{itemize}
\item[(1)] Find the Bethe roots that lie on the line $\mathrm{Re}(x)=\pi$. Denote this set by $A_1$ and its cardinality by $N_1$.
\item[(2)] Find the remaining roots belong to the interval $(\pi/2,\pi)$.
Denote this set by $A_2$ and its cardinality by $N_2$.
\item[(3)] Determine the four roots that are situated just to the left and just to the right of $\pi/2$. We label them $\pp^-_{1}$, $\pp^-_{2}$, $\pp^+_{1}$, $\pp^+_{2}$ with $\pp^-_{2}< \pp^-_{1}\leq \pi/2\leq \pp^+_{1}<\pp^+_{2}$.
\end{itemize} 
The Bethe roots configuration of the ground state, denoted as $\bm{p}_{\rm g}$, takes the following form
\begin{align}
{\textit{case I}}:&\quad \bm{p}_{\rm g}=A_1\cup A_2, \quad \mbox{with}\,\, M=N_1+N_2, \label{case1}\\
{\textit{case II}}:&\quad \bm{p}_{\rm g}=(A_1\cup A_2)\cup\{\pp^-_{1}\},\quad \mbox{with}\,\,M=N_1+N_2+1\,\,\mbox{and}\,\,\epsilon(\pp^-_{1})<-\epsilon(\pp^+_{1}), \label{case2}\\
{\textit{case III}}:&\quad \bm{p}_{\rm g}=(A_1\cup A_2)\setminus\{\pp^+_{1}\},\quad \mbox{with}\,\,M=N_1+N_2-1\,\,\mbox{and}\,\,\epsilon(\pp^-_{1})>-\epsilon(\pp^+_{1}). \label{case3}
\end{align}

It should be noted that the Bethe Ansatz equations \eqref{BAEs(alter)} are independent of $\theta_{\pm}$. Consequently, we can adjust $\theta_{\pm}$ so that $M$ has either the same or different parity from $N_1+N_2$.

\subsection{Elementary excitation}

Due to the parity-restriction of the number of the Bethe roots, The elementary excitation has the following possibilities: (1) replacing one root in the set $\bm{p}_{\rm g}$ with another root outside the set; (2) adding two roots to the set $\bm{p}_{\rm g}$; (3) removing two roots from the set $\bm{p}_{\rm g}$.

Let us first consider \textit{case I} in Eq. \eqref{case1}, the energies for all the possible elementary excitations can be derived as follows
\begin{align}\label{Elementary excitations 1}
\Delta E=\begin{cases} \epsilon(\pp_1^-)-\epsilon(\pp_1^+), & \mbox{replace $\pp^+_1$ in $\bm{p}_{\rm g}$ with $\pp^-_1$}, \\
-\epsilon(\pp_1^+)-\epsilon(\pp_2^+), & \mbox{remove two roots  $\{\pp^+_2,\pp_1^+\}$ from $\bm{p}_{\rm g}$}, \\
\epsilon(\pp_1^-)+\epsilon(\pp_2^-), & \mbox{add two roots $\{\pp^-_2,\pp_1^-\}$ to $\bm{p}_{\rm g}$}.
\end{cases}
\end{align}
Then, we consider \textit{case II} in Eq. \eqref{case2}. We can obtain the energies for all the possible elementary excitations
\begin{align}\label{Elementary excitations 2}
\Delta E=\begin{cases} \epsilon(\pp_2^-)-\epsilon(\pp_1^-), & \mbox{replace $\pp^-_1$ in $\bm{p}_{\rm g}$ with $\pp^-_2$}, \\
-\epsilon(\pp_1^+)-\epsilon(\pp_1^-), & \mbox{remove two roots  $\{\pp^-_1,\pp^+_1\}$ from $\bm{p}_{\rm g}$}.
\end{cases}
\end{align}
For \textit{case III} in Eq. \eqref{case3}, the energies for all the possible elementary excitations read
\begin{align}\label{Elementary excitations 3}
\Delta E=\begin{cases} \epsilon(\pp_1^+)-\epsilon(\pp_2^+), & \mbox{replace $\pp^+_2$ in $\bm{p}_{\rm g}$ with $\pp^+_1$}, \\
\epsilon(\pp_1^+)+\epsilon(\pp_1^-), & \mbox{add two roots  $\{\pp^-_1,\pp^+_1\}$ to $\bm{p}_{\rm g}$}.
\end{cases}
\end{align}

Without losing generality, we study the case where the number of lattice sites is a multiple of 4, i.e.,
$N=4n, n \in \mathbb{N}^+$. Based on the configuration of Bethe roots in the ground and first excited states, one can obtain the ``phase'' diagram in Fig. \ref{Fig: N=1234}(a), which is separated into four regimes by the curve $w_-w_+=1$ and the line $w_-+w_+=0$. In the following, we will discuss the properties of different regions.

\paragraph{Region a}
Once $M$ takes odd values, the Bethe roots configuration corresponding to the ground state is given by $\bm{p}_{\rm g}=A_1 \cup A_2$ with $|\bm{p}_{\rm g}|=\frac{N}{2}+1$. In this region, we find $\epsilon(\pp_1^+)+\epsilon(\pp_2^-)>0$ and $\epsilon(\pp_1^-)+\epsilon(\pp_2^+)<0$. The Bethe root configuration corresponding to the first excited state is therefore given by  $\bm{p}_{\rm e}=(A_1 \cup A_2) \cup \{\pp_1^- \}\setminus \{\pp^+_1\}$ with $|\bm{p}_{\rm e}|=\frac{N}{2}+1$. The elementary excitation energy reads $\epsilon(\pp_1^-)-\epsilon(\pp_1^+)$.

When $M$ is even, the distribution of the ground-state Bethe roots corresponds to two distinct cases:

(1) region a1: we find $\pp_1^++\pp_1^->\pi$, and the Bethe root configuration corresponding to the ground state is given by $\bm{p}_{\rm g}=(A_1 \cup A_2)\cup \{\pp_1^-\}$ with $|\bm{p}_{\rm g}|=\frac{N}{2}+2$. For the elementary excitation, two possible cases may occur \eqref{Elementary excitations 2}, because $\epsilon(\pp_1^+)+\epsilon(\pp_2^-)>0$, the root configuration corresponding to the first excited state is given by $\bm{p}_{\rm e}=(A_1 \cup A_2) \setminus \{\pp^+_1\}$ with $|\bm{p}_{\rm e}|=\frac{N}{2}$. The elementary excitation energy is $-\epsilon(\pp_1^+)-\epsilon(\pp_1^-)$ (see Fig. \ref{region a}(a) for an example).

(2) region a2: we find $\pp_1^++\pp_1^-<\pi$, and the Bethe root configuration corresponding to the ground state is given by $\bm{p}_{\rm g}=(A_1 \cup A_2)\setminus \{\pp^+_1\}$ with $|\bm{p}_{\rm g}|=\frac{N}{2}$. For the elementary excitation, two possible cases may occur \eqref{Elementary excitations 3}, because $\epsilon(\pp_1^-)+\epsilon(\pp_2^+)<0$, the root configuration corresponding to the first excited state is given by $\bm{p}_{\rm e}=(A_1 \cup A_2) \cup \{\pp_1^-\}$ with $|\bm{p}_{\rm e}|=\frac{N}{2}+2$. The elementary excitation energy is $\epsilon(\pp_1^-)+\epsilon(\pp_1^+)$ (see Fig. \ref{region a}(b) for an example).

\begin{center}
\refstepcounter{figure}
\includegraphics[width=6.5cm]{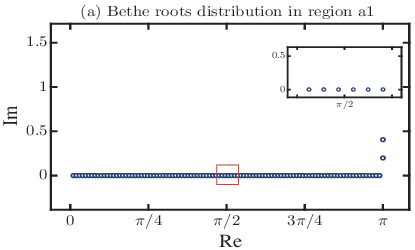}\quad
\includegraphics[width=6.5cm]{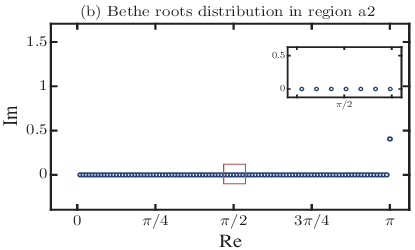}\\[8pt]
\parbox[c]{15.0cm}{\footnotesize{\bf Fig.~5.} 
(a): Distribution of the Bethe roots for $N=100$, $w_+=-1.50$, $w_-=-1.22$, $g=0.04$. Here, $\pp_2^-=1.5270$, $\pp_1^-=1.5582$, $\pp_1^+=1.5893$, $\pp_2^+=1.6205$, the single quasi-particle energy is $\epsilon(\pp_2^-)=0.1752$, $\epsilon(\pp_1^-)=0.0506$, $\epsilon(\pp_1^+)=-0.0741$, $\epsilon(\pp_2^+)=-0.1987$.  When $M$ is odd, the ground state energy is $E(\bm{p}_{\rm g})=-129.0713$. The elementary excitation has three probabilities \eqref{Elementary excitations 1}:  (1): replace $\pp_1^+$ in $\bm{p}_{\rm g}$ with $\pp_1^-$, $\Delta E =\epsilon(\pp_1^-)- \epsilon(\pp_1^+)=0.1247$; (2): remove two roots $\{\pp_2^+,\pp_1^+\}$ from $\bm{p}_{\rm g}$, $\Delta E=-\epsilon(\pp_1^+)- \epsilon(\pp_2^+)=0.2728$; (3): add two roots$\{\pp_1^-,\pp_2^- \}$ to $\bm{p}_{\rm g}$, $\Delta E =\epsilon(\pp_1^-)+ \epsilon(\pp_2^-)=0.2258$. We see that replacing a single root yields a lower excitation energy than adding two roots. When $M$ is even, the ground state energy is $E(\bm{p}_{\rm g})=-129.0207$.  The elementary excitation has two probabilities \eqref{Elementary excitations 2}: (1): replace $\pp_1^-$ in $\bm{p}_{\rm g}$ with $\pp_2^-$, $\Delta E=\epsilon(\pp_2^-)- \epsilon(\pp_1^-)=0.1246$; (2): remove another two roots $\{\pp_1^-,\pp_1^+ \}$ to $\bm{p}_{\rm g}$, $\Delta E=-\epsilon(\pp_1^-)- \epsilon(\pp_1^+) =0.0235$.  Clearly, removing two roots constitutes the elementary excitation.
(b): Distribution of the Bethe roots for $N=100$, $w_+=-1.50$, $w_-=0.70$, $g=0.04$. Here, $\pp_2^-=1.5123$, $\pp_1^-=1.5434$, $\pp_1^+=1.5745$, $\pp_2^+=1.6056$, the energy corresponding to a single root is $\epsilon(\pp_2^-)=0.2338$, $\epsilon(\pp_1^-)=0.1096$, $\epsilon(\pp_1^+)=-0.0148$, $\epsilon(\pp_2^+)=-0.1392$. When $M$ is odd, the ground state energy is $E(\bm{p}_{\rm g})=-128.7423$. The elementary excitation has three probabilities \eqref{Elementary excitations 1}: (1): replace $\pp_1^+$ in $\bm{p}_{\rm g}$ with $\pp_1^-$, $\Delta E =\epsilon(\pp_1^-)- \epsilon(\pp_1^+)=0.1243$; (2): remove two roots $\{\pp_2^+,\pp_1^+\}$ from $\bm{p}_{\rm g}$, $\Delta E=-\epsilon(\pp_1^+)- \epsilon(\pp_2^+)=0.1540$; (3): add two roots$\{\pp_1^-,\pp_2^- \}$ to $\bm{p}_{\rm g}$, $\Delta E =\epsilon(\pp_1^-)+ \epsilon(\pp_2^-)=0.3433$. We see that replacing a single root yields a lower excitation energy than adding two roots. When $M$ is even, the ground state energy is $E(\bm{p}_{\rm g})=-128.7276$.  The elementary excitation has two probabilities \eqref{Elementary excitations 3}: (1): 
replace $\pp_2^+$ in $\bm{p}_{\rm g}$ with $\pp_1^+$, $\Delta E =\epsilon(\pp_1^+)-\epsilon(\pp_2^+)=0.1244$;  (2): add another two roots $\{\pp_1^-,\pp_1^+ \}$ to $\bm{p}_{\rm g}$, $\Delta E =\epsilon(\pp_1^-)+ \epsilon(\pp_1^+)=0.0948$. Obviously, adding two roots constitutes the elementary excitation.
}
\label{region a}
\end{center}

\paragraph{Region b}
We see that the auxiliary function $f(u)$ possesses the following symmetry
\begin{align}
f(\er^{\ir x})|_{w_\pm \to -w_\pm}= -\er^{(2N+4)\ir x} f(\er^{\ir\pi-\ir x}).
\end{align}
Therefore, the Bethe root configurations for all energy levels in region b can be derived directly from those in region a. For the ground and first excited states, one can draw the following conclusions
\begin{itemize}
    \item[(1)]  Once $M$ takes even values, the ground state Bethe root configuration is $\bm{p}_{\rm g}=A_1 \cup A_2$, and that for the first excited state is  $\bm{p}_{\rm e}=(A_1 \cup A_2) \cup \{\pp_1^-\} \setminus\{ \pp^+_1\}$.
    \item[(2)] For odd $M$, in region b1, the Bethe root configurations of the ground state is $\bm{p}_{\rm g}=(A_1 \cup A_2) \cup \{\pp_1^-\}$, while that of the first excited state is $\bm{p}_{\rm e}=(A_1 \cup A_2) \setminus \{\pp_1^+ \}$. In region b2, they are interchanged.
\end{itemize}

We see that the elementary excitation pattern varies with the parameters $w_\pm$ and the integer $M$. Nevertheless, the elementary excitation energy consists of two bare quasi-particle energies (where both removing and adding a root can be considered as a quasi-particle).

In the aforementioned analysis, we have considered only the case where $N = 4n$. For the model with other system lengths, the Bethe root configurations of the ground and first excited states can be analyzed in the same way. The ``phase'' diagrams for different lattice sizes are shown in Fig. \ref{Fig: N=1234}, and the cardinality of the set $A_1 \cup A_2$ in different regions and for different lattice sizes is summarized in Table~\ref{tab:region}.

\begin{center}
\refstepcounter{figure}
\includegraphics[width=6.5cm]{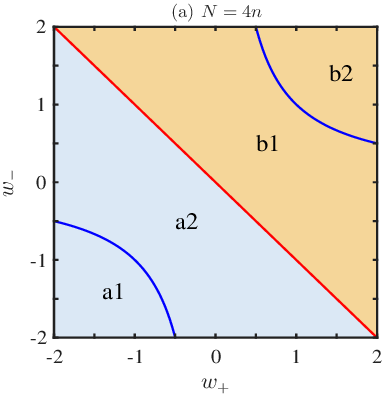}\qquad
\includegraphics[width=6.5cm]{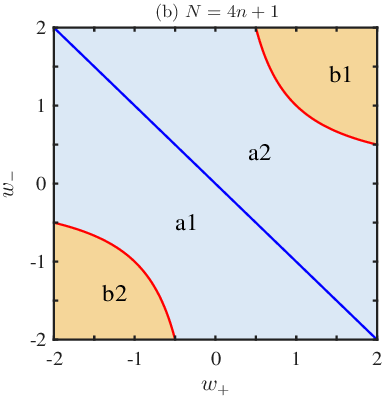}\qquad
\includegraphics[width=6.5cm]{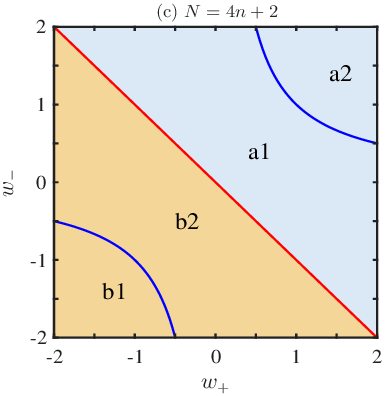}\qquad
\includegraphics[width=6.5cm]{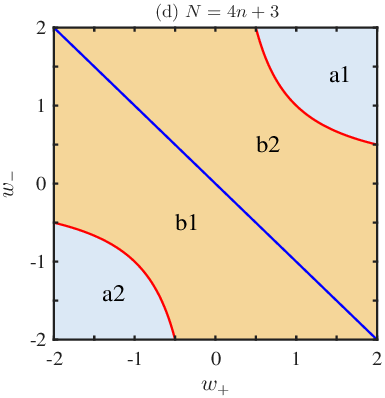}\\[8pt]
\parbox[c]{15.0cm}{\footnotesize{\bf Fig.~6.}
 ``Phase'' diagram of the distribution of the $M$ Bethe roots corresponding to the ground and first excited state for different lattice sizes. 
 In the four panels, regions a1, a2, b1, and b2 each exhibit distinct root distribution patterns for the ground and first excited states. The patterns vary across regions but remain consistent within each region. For a given region, the primary difference for different lattice sizes lies in the total number of Bethe roots. A summation is made in Table \ref{tab:region}. }
 \label{Fig: N=1234}
\end{center}

\begin{center}
\refstepcounter{table}
{\footnotesize{\bf Table 2.} Summation of the cardinality of the set $A_1 \cup A_2$ versus the system size and the regions. 
\vspace{2mm}
\begin{tabular}{|c|c|c|c|c|c|}
    \hline
    \multicolumn{2}{|c|}{\diagbox{Region}{$N_1+N_2$}{$N$} }& $4n$ & $4n+1$ & $4n+2$ & $4n+3$ \\
    \hline
    \multirow{2}{*}{$a$} & a1 & \multirow{2}{*}{$(N+2)/2$} & \multirow{2}{*}{$(N+1)/2$} & \multirow{2}{*}{$N/2$} & $(N-1)/2$ \\
    \cline{2-2}\cline{6-6}
    & a2 & & & & $(N+3)/2$ \\
    \hline
    \multirow{2}{*}{$b$} & b1 & \multirow{2}{*}{$N/2$} & $(N-1)/2$ & \multirow{2}{*}{$(N+2)/2$} & \multirow{2}{*}{$(N+1)/2$} \\
    \cline{2-2}\cline{4-4}
    & b2 & & $(N+3)/2$ & & \\
    \hline
  \end{tabular}}
  \label{tab:region}
\end{center}

\subsection{Degeneracy of the ground and first excited states}
In generic cases, both the ground state and the first excited state are non-degenerate. Nevertheless, this situation changes under specific conditions, namely on the dividing lines between different regions. Let us focus on the case with an even $N$ and study the degeneracy of the ground and first excited states.

Suppose that $u=\er^{x_0}$ is a zero of $f(u)$, which gives  
\begin{eqnarray}\label{el}
\er^{2N x_0}=\frac{(\er^{2x_0}g +1)(1-w_+\er^{x_0})(1-w_-\er^{x_0})}{(g +\er^{2x_0}) (w_+-\er^{x_0})(w_--\er^{x_0})}.
\end{eqnarray}
It thus can be derived that 
\begin{eqnarray}
f(\er^{{\rm i} \pi-x_0})=-\frac{2\er^{-3 x_0} (w_-+w_+) (w_- w_+-1) (\er^{2 x_0}-1) (g+\er^{2x_0})}{(w_- \er^{x_0}-1) (w_+ \er^{x_0}-1)}.
\end{eqnarray}
Consequently, when $(w_-+w_+) (w_- w_+-1)=0$, the zeros of $f(\er^{\ir x})$ exhibit symmetry about $x = \pi/2$; specifically, whenever $x_0$ is a zero, $\pi - x_0$ is a zero as well.

The value of $f(\er^{{\rm i} \pi/2})$ can be calculated directly
\begin{eqnarray}
f(\er^{{\rm i} \pi/2})=
\begin{cases}\label{even odd}
-2 {\rm i} (g-1) (w_-+w_+), & \mbox{even $N$},\\
-2 (g-1) (w_- w_+-1), & \mbox{odd $N$}.
\end{cases}
\end{eqnarray}

\paragraph{Degeneracy on the curve $w_-w_+=1$}
 Under the condition $w_-w_+=1$, the zeros of $f(\er^{\ir x})$ exhibit symmetry about $x = \pi/2$ and $\pi/2$ itself is not a zero. Once $M=N_1+N_2+2k$, both the ground and the first excited state are not degenerate. In the case $M=N_1+N_2+2k+1$, the ground state and first excited state both exhibit a two-fold degeneracy (see Fig. \ref{Degenerate image}(a) for an example)
\begin{align}
&E(\bm{p}_{{\rm g},1})=E(\bm{p}_{{\rm g},2}),\quad \bm{p}_{{\rm g},1}=(A_1 \cup A_2) \setminus \{ \pp_1^+ \}, \quad \bm{p}_{{\rm g},2}=(A_1 \cup A_2) \cup \{ \pp_1^-\},\\
&E(\bm{p}_{{\rm e},1})=E(\bm{p}_{{\rm e},2}),\quad \bm{p}_{{\rm e},1}=(A_1 \cup A_2) \setminus \{ \pp_2^+ \}, \quad \bm{p}_{{\rm e},2}=(A_1 \cup A_2) \cup \{\pp_2^-\}.
\end{align}

\paragraph{Degeneracy on the line $w_-+w_+=0$}
Under the condition $w_-+w_+=0$, the zeros of $f(\er^{\ir x})$ exhibit symmetry about $x = \pi/2$ and $\pi/2$ itself is a zero. Setting $\pp^-_1=\pi/2$, one readily obtains $\pp^-_{2}=\pi -\pp^+_1$. Once $M=N_1+N_2+2k$, the ground state is not degenerate with $E_{\rm g}=E(\bm{p}_{\rm g}),\,\,\bm{p}_{\rm g}=A_1 \cup A_2$, whereas the first excited state exhibits a two-fold degeneracy as follows
\begin{align}
E(\bm{p}_{{\rm e},1})=E(\bm{p}_{{\rm e},2}),\quad \bm{p}_{{\rm e},1}=(A_1 \cup A_2 )\cup \{ \pp_1^-\} \setminus \{\pp_1^+\}, \quad \bm{p}_{{\rm e},2}= (A_1 \cup A_2) \cup \{\pp_1^- ,\pp_2^-\}. 
\end{align}
In the case $M=N_1+N_2+2k+1$, similarly, the ground state is not degenerate with $\bm{p}_{\rm g}=(A_1 \cup A_2) \cup \{\pp_1^- \}$, whereas the first excited state exhibits a two-fold degeneracy (see Fig. \ref{Degenerate image}(b) for an example)
\begin{align}
E(\bm{p}_{{\rm e},1})=E(\bm{p}_{{\rm e},2}),\quad
\bm{p}_{{\rm e},1}=(A_1\cup A_2)\cup \{\pp_2^-\},\quad\bm{p}_{{\rm e},2}=(A_1\cup A_2)\setminus\{\pp_1^+\}.
\end{align}

\begin{center}
\refstepcounter{figure}
\includegraphics[width=6.5cm]{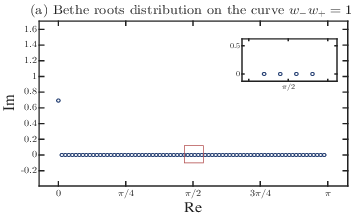}\quad
\includegraphics[width=6.5cm]{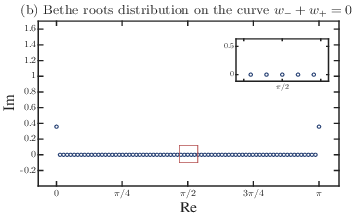}\\[8pt]
\parbox[c]{15.0cm}{\footnotesize{\bf Fig.~7.}
  (a): Bethe roots distributions for the parameters $N=76$, $w_+=2.00$, $w_-=0.50$, $g=0.04$. Here, $\pp_2^-=1.5097$, $\pp_1^-=1.5504$, $\pp_1^+=1.5912$, $\pp_2^+=1.6319$, the energy corresponding to a single root is $\epsilon(\pp_2^-)=0.2444$, $\epsilon(\pp_1^-)=0.0815$, $\epsilon(\pp_1^+)=-0.0815$, $\epsilon(\pp_2^+)=-0.2444$. When $M=N_1+N_2+2k+1$, the ground state energy is $E(\bm{p}_{{\rm g},1})=E(\bm{p}_{{\rm g},2})=-98.4305$ and the first excited state energy is $E(\bm{p}_{{\rm e},1})=E(\bm{p}_{{\rm e},2})=-98.2676$. When $M=N_1+N_2+2k$, the ground state energy is $E(\bm{p}_{\rm g})=-98.5120$ and the first excited state energy is $E(\bm{p}_{\rm e})=-98.3490$. (b): Bethe roots distributions for the parameters $N=76$, $w_+=-1.43$, $w_-=1.43$, $g=0.04$. Here, $\pp_2^-=1.5297$, $\pp_1^-=\pi/2$,  $\pp_1^+=1.6117$, $\pp_2^+=1.6527$, the energy corresponding to a single root is $\epsilon(\pp_2^-)=0.1637$, $\epsilon(\pp_1^-)=0$, $\epsilon(\pp_1^+)=-0.1637$, $\epsilon(\pp_2^+)=-0.3271$. When $M=N_1+N_2+2k$, the ground state energy is $E(\bm{p}_{\rm g})=-98.5956$. The first excited state is given by: $E(\bm{p}_{{\rm e},1})=E(\bm{p}_{{\rm e},2})=-98.4319$. When $M=N_1+N_2+2k+1$, the ground state energy is $E(\bm{p}_{\rm g})=-98.5956$. The first excited state is given by: $E(\bm{p}_{{\rm e},1})=E(\bm{p}_{{\rm e},2})=-98.4319$.}
 \label{Degenerate image}
\end{center}

\section{Ground state energy in the thermodynamic limit}\label{sec:therm}

For a finite-size system, it is found that the elementary excitations mainly depend on the two roots $\pp_1^+$ and $\pp_1^-$. When $N$ becomes large, these roots approach $\pi/2$ with a speed of $1/N$ (see Fig.~\ref{Fig: 1/N}):
\begin{align}
    \pp_1^\pm(N)-\frac{\pi}{2}\sim \frac{1}{N}.
\end{align}
Consequently, the elementary excitation energy is also of order $1/N$. In the thermodynamic limit, the gap between the ground state and the first excited state vanishes. The set $A_1 \cup A_2$ will always yield the ground state energy up to an error of order $1/N$. In the following, we derive the analytic expression for the ground-state energy in the thermodynamic limit using the distribution of Bethe roots $\{\mu_j\}$. 
\begin{center}
\refstepcounter{figure}
\includegraphics[width=0.48\textwidth]{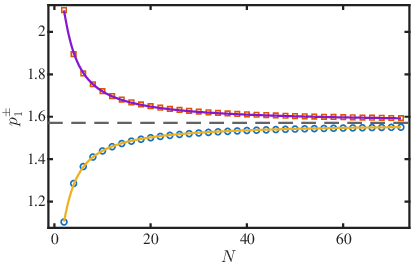}
\parbox[c]{15.0cm}{\footnotesize{\bf Fig.~8.}
The curves of $\pp_1^\pm$ versus the system size $N$, with fixed boundary parameters $w_+=1.74$, $w_-=0.45$, and $g=0.04$. Here the blue circles and red squares denote the numerical data of $\pp_1^-$ and $\pp_1^+$, respectively, while the yellow and purple curves correspond to the fitting functions $p_{1}^{-}( N ) =\pi/{2}-1.41 N^{-1}+0.95 N^{-2}$ and $p_{1}^{+}( N ) =\pi/{2}+1.60 N^{-1}-1.07 N^{-2}$, respectively. The gray dashed line in the middle represents $\pi/2$.} 
 \label{Fig: 1/N}
\end{center}

First, recall the Bethe Ansatz equations \eqref{BAEs}
\begin{align}\label{BAE:2}
&\left[\frac{\sinh(\mu_j+\frac{\ir\pi}{4})}{\sinh(\mu_j-\frac{\ir\pi}{4})}\right]^{2N}\frac{\sinh(\mu_j-\ir\alpha_--\frac{\ir\pi}{4}) \sinh(\mu_j-\ir\alpha_+-\frac{\ir\pi}{4}) }{\sinh(\mu_j+\ir\alpha_-+\frac{\ir\pi}{4}) \sinh(\mu_j+\ir\alpha_++\frac{\ir\pi}{4}) }\nonumber\\
&\times \frac{\cosh(\mu_j-\beta-\frac{\ir\pi}{4}) \cosh(\mu_j+\beta-\frac{\ir\pi}{4})}{\cosh(\mu_j-\beta+\frac{\ir\pi}{4}) \cosh(\mu_j+\beta+\frac{\ir\pi}{4})}=1,\quad j=1,\ldots,M.
\end{align}
Under the substitution $\alpha_k \to \alpha_k + \pi$, $\theta_k \to \theta_k + \pi$, the constraint in  \eqref{Constraint:XX}, the BAEs \eqref{BAEs} and the energy expression \eqref{Energy expressions 1} still hold. As a consequence, the spectrum of the system is invariant. Therefore, we can choose $0 < \alpha_\pm< \pi$.

Most of the Bethe roots $\{\mu_j\}$ distribute on two lines 
\begin{align}
&({\rm i}):\,\,{\rm Im}(\mu_j)=0, \quad -4\leq \varepsilon(\mu_j)\leq 0,\\
&({\rm ii}):\,\,{\rm Im}(\mu_j)=\frac{\pi}{2},\quad 0\leq \varepsilon(\mu_j)\leq 4,\quad \varepsilon(u)=-\frac{4}{\cosh(2u)}.
\end{align}
It should be noted that $\mu_j=0,\ir\pi/2$ are trivial solutions. 
In case of $|\cot\alpha_k|>1$ ($\alpha_k\in (0,\frac{\pi}{4})\cup (\frac{3\pi}{4},\pi)$), there exists the following purely imaginary solutions
\begin{align}
\mu_j&\to\ir\alpha_j+\frac{\ir\pi}{4},\quad |\cot\alpha_k|>1,\no\\
\varepsilon(\mu_j)&=2\cot(\alpha_k)+2\tan(\alpha_k)+O(\tfrac1N)=\frac{4}{\sin(2\alpha_k)}+O(\tfrac1N).
\end{align}
We see that these purely imaginary roots should be included in the ground state once $\alpha_k\in(3\pi/4,\pi)$. 

Taking the logarithm of both sides of \eqref{BAE:2}, we arrive at
\begin{align}
I_j=2N\xi(0,\mu_j)-\xi(\alpha_-,\mu_j)-
\xi(\alpha_+,\mu_j)-\bar\xi(\beta,\mu_j)-\bar\xi(-\beta,\mu_j),
\end{align}
where 
\begin{align}
\xi(n,u)&=\frac{1}{2\ir\pi}\ln\left[ \frac{\sinh(u-\ir n-\frac{\ir \pi}{4})}{\sinh(u+\ir n+\frac{\ir \pi}{4})}\right],\\
\bar\xi(n,u)&=\frac{1}{2\ir\pi}\ln \left[\frac{\cosh(u-n-\frac{\ir \pi}{4})}{\cosh(u+n+\frac{\ir \pi}{4})}\right],
\end{align}
and $I_j$ is an integer.
In the thermodynamic limit, the Bethe roots for the ground state accumulate on the real axis and fill the interval $[-W, W]$ densely. Their distribution is described by a density function $\rho(\mu),\,u\in\mathbb{R}$, which satisfies
\begin{align}
\rho(\mu)=\kappa(0,\mu)-\frac{1}{2N}\kappa(\alpha_-,\mu)-\frac{1}{2N}
\kappa(\alpha_+,\mu)-\frac{1}{2N}\bar\kappa(\beta,\mu)-\frac{1}{2N}\bar\kappa(-\beta,\mu_j)-\frac{1}{2N}\delta(0),
\end{align}
where 
\begin{align}
\kappa(n,u)&=\frac{\dr \xi(n,u)}{\dr u}=\frac{1}{\pi}\frac{\cos (2n)}{\cosh (2u)+\sin (2n)},\\
\bar\kappa(n,u)&=\frac{\dr\bar\xi(n,u)}{\dr u}=-\frac{1}{\pi}\frac{\cosh (2n)}{\cosh (2 u)+\ir\sinh (2n)}.
\end{align}
In the interval $[-W,W]$, the total number of real roots should be $M_{\rm r}$, and satisfy 
\begin{align}
2M_{\rm r}&=2N\int_{-W}^{W}\rho(u)\dr u.
\end{align}
Given that $M_{\mathrm{r}}\sim \frac{N}{2}$ (see Table~\ref{tab:region}) and $\int_{-\infty}^{\infty}\kappa(0,u)\dr u=\frac12$, it follows that $W$ goes to infinity in the thermodynamic limit.
Since roots $\mu_j\in \mathbb{R}$ with large $|\mu_j|$ make only a small contribution to the energy, we can neglect $O(1/N)$ terms and take the limit $W\to\infty$.

Let us introduce the following useful identities 
\begin{align}
&-\int_{-\infty}^{\infty}\kappa(a,u)\varepsilon(u) \dr u=
\begin{cases}
\frac{8a\pi^{-1}+2 \cos (2a)-2}{\sin (2 a)}, & a\in(0,3\pi/4), \\
\frac{8 a\pi^{-1}+2\cos (2a)-10}{\sin (2a)}, & a\in(3\pi/4, \pi),
\end{cases}\\
&-\int_{-\infty}^{\infty}[\bar\kappa(a,u)+\bar \kappa(-a+u)]\varepsilon(u) \dr u=-\frac{16a}{\pi  \sinh (2a)},\quad a\in\mathbb{R}.
\end{align}

\paragraph{(i): $\alpha_+,\alpha_-\in(0,3\pi/4)$} In this case, all the roots for the ground state should lie on the real axis, and the energy reads
\begin{align}\label{TBA: the ground state energy 1}
E_{\rm g}&=E_0+N\int_{-\infty}^{\infty}\varepsilon(u)\rho(u) \dr u\no\\
&=-\frac{4N}{\pi}+\frac{4\alpha_-\pi^{-1}+\cos (2 \alpha_-)-1}{\sin (2\alpha_-)}+\frac{4\alpha_+\pi^{-1}+\cos (2 \alpha_+)-1}{\sin (2\alpha_+)}-\frac{8\beta}{\pi  \sinh (2\beta)}\no\\
&\quad +2-\cot\alpha_--\cot\alpha_+\no\\
&=-\frac{4N}{\pi}+\frac{4\alpha_--2\pi}{\pi\sin (2\alpha_-)}+\frac{4\alpha_+-2\pi}{\pi\sin (2\alpha_+)}-\frac{8\beta}{\pi  \sinh (2\beta)}+2.
\end{align}
\paragraph{(ii): $\alpha_+\in(0,3\pi/4)$, $\alpha_-\in(3\pi/4,\pi)$} In this case, one root lies on the imaginary axis and the remaining roots lie on the real axis. The ground state energy reads
\begin{align}
E_{\rm g}&=E_0+N\int_{-\infty}^{\infty}\varepsilon(u)\rho(u) \dr u+\frac{4}{\sin(2\alpha_-)}\no\\
&=-\frac{4N}{\pi}+\frac{4\alpha_-\pi^{-1}+\cos (2 \alpha_-)-5}{\sin (2\alpha_-)}+\frac{4\alpha_+\pi^{-1}+\cos (2 \alpha_+)-1}{\sin (2\alpha_+)}-\frac{8\beta}{\pi  \sinh (2\beta)}\no\\
&\quad +2-\cot\alpha_--\cot\alpha_++\frac{4}{\sin(2\alpha_-)}\no\\
&=-\frac{4N}{\pi}+\frac{4\alpha_--2\pi}{\pi\sin (2\alpha_-)}+\frac{4\alpha_+-2\pi}{\pi\sin (2\alpha_+)}-\frac{8\beta}{\pi  \sinh (2\beta)}+2.
\end{align}
\paragraph{(iii): $\alpha_+,\alpha_-\in(3\pi/4,\pi)$} In this case, two roots lie on the imaginary axis and the remaining roots lie on the real axis. The ground state energy is
\begin{align}\label{TBA: the ground state energy 3}
E_{\rm g}&=E_0+N\int_{-\infty}^{\infty}\varepsilon(u)\rho(u) \dr u+\frac{4}{\sin(2\alpha_-)}+\frac{4}{\sin(2\alpha_+)}\no\\
&=-\frac{4N}{\pi}+\frac{4\alpha_-\pi^{-1}+\cos (2 \alpha_-)-5}{\sin (2\alpha_-)}+\frac{4\alpha_+\pi^{-1}+\cos (2 \alpha_+)-5}{\sin (2\alpha_+)}-\frac{8\beta}{\pi  \sinh (2\beta)}\no\\
&\quad +2-\cot\alpha_--\cot\alpha_++\frac{4}{\sin(2\alpha_-)}+\frac{4}{\sin(2\alpha_+)}\no\\
&=-\frac{4N}{\pi}+\frac{4\alpha_--2\pi}{\pi\sin (2\alpha_-)}+\frac{4\alpha_+-2\pi}{\pi\sin (2\alpha_+)}-\frac{8\beta}{\pi  \sinh (2\beta)}+2.
\end{align}
We see that the isolated imaginary Bethe roots contribute nothing to the ground state energy. In the limit $\beta\to \infty$, Eq. \eqref{TBA: the ground state energy 3} also gives the analytical expression for the ground state energy of the open XX spin chain with diagonal boundaries in the thermodynamic limit. The calculated results can be cross-checked with Refs. \cite{Qiao:2021,Li:2014,murgan:2006boundary}.

\section{Exact spectrum of the open XX model with diagonal boundary fields}\label{sec:diag}

\subsection{Reduction of the Bethe Ansatz equations}

In the limit $\beta\to \infty$ $(g\to 1)$, 
the two boundary magnetic fields all point along the $z$-direction with
\begin{eqnarray}
\vec{h}_{\rm l}=(0,\,0,\,w_-),\quad \vec{h}_{\rm r}=(0,\,0,\,w_+).
\end{eqnarray}
In this case, the $U(1)$ symmetry of the system is recovered, and the corresponding BAEs are
\begin{align}
&\tilde f(\er^{\ir p_j})=0,\quad j=1,\ldots,M',\label{BAE:U1}\\
&\tilde f(u)=u^{2N}\prod_{\sigma =w_\pm}{( \sigma-u ) -}\prod_{\sigma =w_\pm}{( 1-u\sigma )}.\label{def:tilde:f}
\end{align}
Here, the integer $M'$ is the magnon number and can take values ranging from 0 to $N$.
Then, there is a question: can we retrieve the anti-mentioned BAEs \eqref{BAE:U1} from our original one \eqref{BAEs(alter)} in the limit $g\to 1$?

We see that
\begin{align}
&\lim_{g\to 1}f(u) =(u^2+1)\tilde{f}(u).
\end{align}
When $g\to 1$, two zeros of $f(u)$ will go to $\pm\ir$, and the other zeros are given by the equation $\tilde{f}(u)=0$.
A physical solution of BAE \eqref{Energy expressions} is obtained by selecting $M$ Bethe roots from the $N+1$ non-trivial interdependent zero roots of $f(\er^{{\rm i} u})$. Depending on whether the root $p=\pi/2$ is included in the selected set, the configurations of Bethe roots can be  divided into two classes:
\begin{itemize}
    \item[(1)] The root $p = \pi/2$ is included, which does not contribute to the energy ($\epsilon(\pi/2) = 0$). The remaining $M-1$ “effective” roots satisfy $\tilde f(\er^{\ir p}) = 0$ with $\epsilon(p) \neq 0$.
    \item[(2)] The root $p = \pi/2$ is not included, and all $M$ roots in the set satisfy $\tilde f(\er^{\ir p}) = 0$.
\end{itemize}

Due to the constraint in Eq. \eqref{Constraint:XX}, the number of Bethe roots $M$ in Eq. \eqref{BAE:U1} must exhibit a specific parity. However, in the limit $g \rightarrow 1$, one Bethe root may approach $\pi/2$ 
and become energy-ineffective ($\epsilon(\pi/2)=0$). Consequently, the total number of remaining effective Bethe roots, denoted $M_{\rm e}$, that satisfy $\tilde f(\er^{\ir p})=0$ is either $M$ or $M-1$. In this case, $M_{\rm e}$ can take integer values from $0$ to $N$, and one can recover the conventional BAEs in Eq. \eqref{BAE:U1}.


\subsection{Distributions of Bethe roots in the ground state and the first excited state}

In the diagonal case, the Bethe roots $\{p_1,...,p_M\}$ should satisfy BAEs \eqref{BAE:U1}, and the number of Bethe roots can range from 0 to $N$. Consequently, the Bethe roots configuration in the ground state is always given by $\bm{p}_{\rm g}=A_1 \cup A_2$. Analogous to the non-diagonal case, we label the four roots located just to the left and just to the right of $\pi/2$ as $\pp^-_1$, $\pp^-_{2}$, $\pp^+_{1}$, $\pp^+_{2}$, with $\pp^-_{2}< \pp^-_{1}\leq \pi/2\leq \pp^+_{1}<\pp^+_{2}$.
The elementary excitation now has two possibilities \cite{lieb1961two,biegel:2004spectrum}: (1) add one root ${\pp_1^-}$ to the set $\bm{p}_{\rm g}$; (2) remove one root ${\pp_1^+}$ from the set $\bm{p}_{\rm g}$.

We consider the case where the number of lattice sites is a multiple of 4 ($N=4n,\,n\in\mathbb{N}^+$). As in the non-diagonal case, the ``phase'' diagram is separated into four regimes by the curve $w_-w_+=1$ and the line $w_-+w_+=0$, see Figure \ref{Fig: N=1234}. The Bethe roots configurations of the first excited state in each region can be derived

(1) regions a1 and b1: we find $\pp_1^-+\pp_1^+ <\pi$, and the root distribution corresponding to the first excited state is thus given by  $\bm{p}_{\rm e}=(A_1\cup A_2)\setminus\{\pp_1^+\}$.

(2) regions a2 and b2: we find $\pp_1^-+\pp_1^+ >\pi$, and the root distribution corresponding to the first excited state is thus given by  $\bm{p}_{\rm e}=(A_1\cup A_2)\cup\{\pp_1^-\}$.


In contrast to the non-diagonal case, the elementary excitation always consists of a single bare quasi-particle.

\paragraph{Degeneracy of the ground and first excited states}
Provided that $(w_+w_--1)(w_++w_-)=0$, the zeros of $\tilde f(\er^{\ir x})$ exhibit symmetry about $x = \pi/2$.
In addition, we can calculate the value of $\tilde f(\er^{\ir x})$ at the point $x=\pi/2$
\begin{align}
\tilde{f}(\er^{{\rm i} \pi/2})=
\begin{cases}
2 (w_-w_+-1), & \mbox{even $N$},\\
2\ir  (w_-+w_+), & \mbox{odd $N$}.
\end{cases}
\end{align}

In the curve $w_+w_-=1$, the Bethe roots in the vicinity of $\pi/2$ are $\pp_1^-=\pi/2$, $\pp_2^-=\pi-\pp_1^+$. This results in a two-fold degeneracy of the ground state
\begin{align}
E(\bm {p}_{{\rm g},1})=E(\bm {p}_{{\rm g},2}),\quad 
\bm {p}_{{\rm g},1}=A_1\cup A_2,\quad \bm {p}_{{\rm g},2}=(A_1\cup A_2)\cup \{\pp_1^-\}.
\end{align}
and a four-fold degeneracy of the first excited state
\begin{equation}
\begin{aligned}
&E(\bm p_{{\rm e},1})=E(\bm p_{{\rm e},2})=E(\bm p_{{\rm e},3})=E(\bm p_{{\rm e},4}),\\
&\bm p_{{\rm e},1}=(A_1\cup A_2)\setminus \{\pp_1^+\},\quad\bm p_{{\rm e},2}=(A_1\cup A_2)\cup \{\pp_1^-,\pp_2^-\},\\
&\bm p_{{\rm e},3}=(A_1\cup A_2)\cup \{\pp_1^-\}  \setminus \{\pp_1^+\},\quad\bm p_{{\rm e},4}=(A_1\cup A_2)\cup \{\pp_2^-\}.
\end{aligned}
\end{equation}
On the line $w_++w_-=0$, the Bethe roots near $\pi/2$ are given by: $\pp_1^-=\pi-\pp_1^+$, $\pp_2^-=\pi-\pp_2^+$.  Consequently, the ground state is non-degenerate, characterized by $\bm p_{\rm g}=A_1\cup A_2$, whereas the first excited state is two-fold degenerate with
\begin{align}
&E(\bm {p}_{{\rm e},1})=E(\bm {p}_{{\rm e},2}),\quad \bm {p}_{{\rm e},1}=(A_1\cup A_2)\setminus \{\pp_1^+\},\quad \bm {p}_{{\rm e},2}=(A_1\cup A_2)\cup\{\pp_1^-\}.
\end{align}

Analogously, we can study the model for other system sizes.


\section{Conclusion}
\label{sec:Conclusion}

In this paper, we study the exact spectrum of the open XX chain with constrained non-diagonal boundary fields. The boundary fields break the $U(1)$ symmetry of the model. Nevertheless, the exact spectrum can still be characterized by the conventional homogeneous Bethe Ansatz equations, where the number of Bethe roots has a fixed parity and can range from 0 to $N+1$.
In the Bethe Ansatz equations, all Bethe roots (or quasi-momenta) are decoupled and correspond to the zeros of an auxiliary function. Analyzing these zeros makes it possible to obtain the complete spectrum of the model in a simple way. 

We study the ground state and the first excited state of the model for finite system sizes, reaching up to hundreds or even thousands sites. The patterns of elementary excitations are demonstrated in different parameter regions. It is shown that the elementary excitation always consists of two bare quasi-particles, in contrast to the case with diagonal boundary fields.
In the thermodynamic limit, the model resides in a gapless phase, and an exact analytic expression for the ground state energy is obtained.

The existence of the conventional Bethe Ansatz equations enables the study of the thermodynamic properties of the model-including entropy, free energy, and specific heat-via the thermodynamic Bethe Ansatz approach, which will be presented elsewhere.

Another natural next step is to study the exact spectrum of the XY chain with constrained non-diagonal boundary fields \cite{Yang:2005tq,Zhang:2022xyz}. As in the XX model, conventional BAEs are expected to exist, and the number of Bethe rapidities again has a fixed parity. In contrast to the XX model, the BAEs for the XY chain must be parameterized by elliptic functions and may involve additional boundary parameters, posing a greater challenge. 

We are also interested in the exact eigenstates of the XX model discussed in this paper. In previous works, the Bethe-type eigenstates (Bethe states) of the XXZ model with constrained non-diagonal boundary fields have been constructed using generalized Bethe Ansatz methods \cite{cao2003exact,zhang2021phantom,Zhang:2021chiral}. However, the normalization of these Bethe states remains challenging. For the XX case, in contrast, the Bethe roots are non-entangled, and the normalization of the Bethe states is therefore expected to be considerably simpler than in the XXZ case.

\appendix
\section{Derivation of the BAEs for the XX spin chain}\label{App:A}

From Ref. \cite{wang2015off}, the exact solution of the open XXZ chain defined in Eq. \eqref{Hamiltonian: XXZ} can be parameterized by the following Bethe Ansatz equations 
\begin{align}
&2c\sinh(2\nu_j)\sinh(2\nu_j+2\eta)\sinh^{2N}(\nu_j+\eta)\sinh^{2N}(\nu_j)\nonumber\\
&+a(\nu_j)Q(\nu_j-\eta) + d(\nu_j)Q(\nu_j+\eta)
=0,\quad j=1,\ldots,N,
\end{align}
where \begin{align}
a(u) &= -2^{2} \frac{ \sinh(2u + 2\eta) }{\sinh(2u + \eta) } 
\prod_{\sigma=\pm}\sinh(u - \bar\alpha_{\sigma}) \cosh(u - \bar\beta_{\sigma})\sinh^{2N}(u+\eta),\\
d(u) &= -2^{2} \frac{ \sinh(2u) }{\sinh(2u + \eta) } 
\prod_{\sigma=\pm}\sinh(u + \bar\alpha_{\sigma}+\eta) \cosh(u+\bar\beta_{\sigma}+\eta)\sinh^{2N}(u),\\
c &=\cosh[ (N+1)\eta + \bar\alpha_{-} + \bar\beta_{-} + \bar\alpha_{+} + \bar\beta_{+} ] - \cosh(\theta_{-} - \theta_{+}),
\end{align}
and the $Q$-function reads
\begin{align}
Q(u)=\prod_{j=1}^{N} \sinh(u - \nu_j) \sinh(u + \nu_j + \eta).
\end{align}
The eigenvalue of the Hamiltonian defined in \eqref{Hamiltonian: XXZ} reads
\begin{align}
E=2\sum_{j=1}^N\frac{\sinh^2\eta}{\sinh\nu_j\sinh(\nu_j+\eta)}+(N-1)\cosh\eta-\sinh\eta(\coth\bar\alpha_++\coth\bar\alpha_-+\tanh\bar\beta_++\tanh\bar\beta_-).
\end{align}
By setting $\eta=\ir\pi/2$, one obtains the following Bethe Ansatz equations (BAEs) for the generic XX model defined in Eq. \eqref{xx Hamiltonian} (without the constraint given in Eq. \eqref{Constraint:XX})
\begin{align}
&-\ir \tanh (2\nu_j) \prod_{\sigma=\pm}\sinh (\nu_j-\ir \alpha_\sigma)\prod_{s=\pm1}\cosh (\nu_j+s\beta ) [\ir \cosh (\nu_j)]^{2N}\prod_{k=1}^N[-\ir \sinh (\nu_j+\nu_k) \cosh (\nu_j-\nu_k)]\nonumber\\
&+\ir \tanh (2\nu_j) \prod_{\sigma=\pm}\cosh (\nu_j+\ir \alpha_\sigma)\prod_{s=\pm1}\sinh(\nu_j+s\beta)[\sinh(\nu_j)]^{2N}\prod_{k=1}^N[-\ir \sinh (\nu_j+\nu_k) \cosh (\nu_j-\nu_k)]\nonumber\\
&+2\left(\cos[(N+1)\pi/2+\alpha_-+\alpha_+]-\cos(\theta_--\theta_+)\right)[\ir \sinh(\nu_j)\cosh (\nu_j)]^{2N+2}=0,\qquad j=1,\ldots,N.\label{BAE:generic:xx}
\end{align}

Under the constraint \eqref{Constraint:XX}, the Bethe roots $\{\nu_{M+1},\nu_{M+2},\ldots,\nu_N\}$ tend to infinity \cite{zhang2021phantom}, leading to the following  Bethe Ansatz equations for the remaining regular Bethe roots $\{\mu_{1},\ldots,\mu_M\}=\{\nu_{1}+\frac{\eta}{2},\ldots,\nu_M+\frac{\eta}{2}\}$
\begin{align}
\frac{a(\mu_j-\frac{\eta}{2})}{d(\mu_j-\frac{\eta}{2})}\prod_{k=1}^M\frac{\sinh(\mu_j-\mu_k-\eta)\sinh(\mu_j+\mu_k-\eta)}{\sinh(\mu_j-\mu_k+\eta)\sinh(\mu_j+\mu_k+\eta)}=-1,\quad j=1,\ldots,M.
\end{align}
In the XX case, we arrive at the BAEs \eqref{BAEs}.

\addcontentsline{toc}{chapter}{Acknowledgment}
\section*{Acknowledgment}
We acknowledge financial support from
the National Natural Science Foundation of China (Grant Nos. 12575007, 12247103, 12105221), and Shaanxi Fundamental Science Research Project for Mathematics and Physics (Grant No. 22JSZ005).

\bibliographystyle{iopart-num}
\bibliography{XX_ref}

\end{document}